\documentclass[aps,showpacs,preprintnumbers,amsmath, amssymb]{revtex4}

\usepackage{float}
\usepackage{graphics,epsfig}
\usepackage{graphicx}
\usepackage{dcolumn}
\usepackage{bm}
\usepackage{mathrsfs}

\begin{document}
\baselineskip=0.8 cm

\title{{\bf The existence of null circular geodesics outside extremal
spherically symmetric asymptotically flat hairy black holes}}
\author{Yan Peng$^{1}$\footnote{yanpengphy@163.com}}
\affiliation{\\$^{1}$ School of Mathematical Sciences, Qufu Normal University, Qufu, Shandong 273165, China}

\vspace*{0.2cm}
\begin{abstract}
\baselineskip=0.6 cm
\begin{center}
{\bf Abstract}
\end{center}

The existence of null circular geodesics has been proved in the background of
non-extremal spherically symmetric asymptotically flat black holes in previous works.
Then it is an interesting question that whether extremal black holes possess null circular geodesics outside horizons.
In the present paper, we pay attentions to the extremal spherically symmetric asymptotically flat hairy black
holes. We show the existence of the fastest trajectory to circle a extremal black hole.
As the fastest trajectory corresponds to the position of null circular geodesics,
we prove that null circular geodesics exist outside extremal spherically symmetric
asymptotically flat hairy black holes. We also point out that our proof also 
works for non-extremal black holes.

\end{abstract}

\maketitle
\newpage
\vspace*{0.2cm}

\section{Introduction}

It is usually believed that highly curved black hole spacetimes may be generally
characterized by the existence of null circular
geodesics outside horizons \cite{c1,c2}.
The null circular geodesics are the way that massless
particles can circle the central black holes.
It provides valuable information on the structure of the spacetime geometry,
which is closely related to various remarkable properties
of black hole spacetimes and has attracted
a lot of attentions \cite{c3,c4,c5}.

The presence of circular null geodesics has many important implications on astronomical observations,
such as the gravitational lensing, the black hole shadow, as well as the gravitational waves \cite{c6,c7,c8,c9,c10,c11,c12,c13,c14,c15,c16}.
The circular null geodesics also give the lower bound of the effective radius of scalar field hairs outside
black holes \cite{s1,s2,s3,s4,s5,s6}. The characteristic resonances of perturbed black holes
can be determined by the instability properties of circular null geodesics \cite{r1,r2,r3,r4,r5,r6,r7,r8}.
And massless fields can pile up on stable circular null geodesics, which may lead to nonlinear instabilities
in the highly curved spacetimes \cite{us1,us2,us3,us4,us5,us6,us7,us8}.
It was also showed that the null circular geodesic provides
the fastest way to circle a central black hole \cite{ST1}.

The existence of null circular geodesics can be predicted by
analyzing the non-linearly coupled Einstein-mater field equations.
In non-extremal hairy black-hole spacetimes, the existence of null circular
geodesics has been proved in the asymptotically flat background \cite{ST2}.
And null circular geodesics also exist in the stationary axi-symmetric
non-extremal black hole spacetimes \cite{ST3}.
Along this line, Hod raised a physically interesting question that
whether null circular geodesics can also exist
outside extremal black holes \cite{ST4}.
Hod found that the Einstein-matter field equations
seem to fail to provide a proof for the existence
of null circular geodesics in extremal black hole spacetimes.
So it is interesting to further examine whether
null circular geodesics exist outside extremal black holes.

In this work, we firstly introduce the extremal hairy black hole spacetimes.
We prove the existence of null circular geodesics
by analyzing the existence of the circular trajectory
around central extremal hairy black holes.
At last, we summarize our main results in the conclusion section.

\section{The existence of null circular geodesics in extremal black holes}

We study physical and mathematical properties of null circular geodesics in the background of
extremal hairy black hole spacetimes. The spherically symmetric curved line element of four
dimensional asymptotically flat hairy black hole is described by \cite{ss1,ss2}
\begin{eqnarray}\label{AdSBH}
ds^{2}&=&-e^{-2\delta}\mu dt^{2}+\mu^{-1}dr^2+r^2(d\theta^2+\sin^2\theta d\phi^{2}).
\end{eqnarray}
Here $\delta$ and $\mu$ are metric solutions only depending
on the Schwarzschild areal coordinate r. And angular coordinates are
$\theta\in [0,\pi]$ and $\phi\in [0,2\pi]$. The black hole horizon $r_{H}$ is obtained
from $\mu(r_{H})=0$. At the regular horizon, $\delta$ is finite.
At the infinity, there is $\mu(\infty)=1$ and $\delta(\infty)=0$ in the
asymptotically flat spacetimes. We study null circular geodesics in the equatorial plane
with $\theta=\frac{\pi}{2}$.

In the following, we follow the analysis of \cite{s2} to obtain the null circular geodesic equation.
The Lagrangian describing the geodesics is given by
\begin{eqnarray}\label{AdSBH}
2\mathcal{L}=-e^{-2\delta}\mu \dot{t}^2+\mu^{-1}\dot{r}^2+r^2\dot{\phi}^2,
\end{eqnarray}
where the dot represents differentiation with respect to proper time.

The Lagrangian is independent of coordinates t and $\phi$, leading to
two constants of motion labeled as E and L.
From the Lagrangian (2), we obtain the generalized momenta expressed as \cite{c2}
\begin{eqnarray}\label{BHg}
p_{t}=g_{tt}\dot{t}=-e^{-2\delta}\mu \dot{t}=-E=const,
\end{eqnarray}
\begin{eqnarray}\label{BHg}
p_{\phi}=g_{\phi\phi}\dot{\phi}=r^2\dot{\phi}=L=const,
\end{eqnarray}
\begin{eqnarray}\label{BHg}
p_{r}=g_{rr}\dot{r}=\mu^{-1}\dot{r}~.
\end{eqnarray}

The Hamiltonian of the system is
$\mathcal{H}=p_{t}\dot{t}+p_{r}\dot{r}+p_{\phi}\dot{\phi}-\mathcal{L}$,
which implies
\begin{eqnarray}\label{BHg}
2\mathcal{H}=-E\dot{t}+L\dot{\phi}+g_{rr}\dot{r}^2=\delta=const.
\end{eqnarray}
The case of null geodesics corresponds to the value $\delta=0$.

From relations (3), (4) and (6), we find
\begin{eqnarray}\label{BHg}
\dot{r}^2=\frac{1}{g_{rr}}[E\dot{t}-L\dot{\phi}]
\end{eqnarray}
for null geodesics.

Equations (3) and (4) yield relations
\begin{eqnarray}\label{BHg}
\dot{t}=\frac{E}{e^{2\delta}\mu},~~~~~~~~~~~~~~\dot{\phi}=\frac{L}{r^{2}}.
\end{eqnarray}

Substituting Eqs. (8) into (7), one finds
\begin{eqnarray}\label{BHg}
\dot{r}^2=\mu[\frac{E^2}{e^{2\delta}\mu}-\frac{L^2}{r^2}].
\end{eqnarray}

The null circular geodesic equation $\dot{r}^2=(\dot{r}^2)'=0$ can be expressed as
\begin{eqnarray}\label{BHg}
2e^{-2\delta}\mu-r(e^{2\delta}\mu)'=0.
\end{eqnarray}

The null circular geodesics usually correspond to the
circular trajectory with the shortest orbital period \cite{ST1}.
In the following, we would like to search for the circular trajectory
with the shortest orbital period as measured by asymptotic observers.
In order to minimize the orbital period for a given
radius r, one should move as close as possible to the speed of light.
In this case, the orbital period can be obtained from Eq. (1) with
$ds=dr=d\theta=0$ and $\Delta\phi=2\pi$ \cite{ST1,ST2}:
\begin{eqnarray}\label{AdSBH}
T(r)=-\frac{2\pi\sqrt{-g_{tt}g_{\phi\phi}}}{g_{tt}}=\frac{2\pi r}{e^{-\delta}\sqrt{\mu}}.
\end{eqnarray}

The circular trajectory with the shortest orbital period is characterized by
\begin{eqnarray}\label{AdSBH}
T'(r)=0,
\end{eqnarray}
where the solution is the radius of the fast circular trajectory.

The condition (12) yields the fast circular trajectory equation
\begin{eqnarray}\label{BHg}
2e^{-2\delta}\mu-r(e^{2\delta}\mu)'=0.
\end{eqnarray}

We find that the null circular geodesic equation (10)
is identical to the fastest circular trajectory equation (13).
So the extreme period circle radius equation and
the null circular geodesics equation share the same roots.
If we can prove the existence of the fastest circular trajectory,
the null circular geodesic exists outside black holes.
In fact, there are relations $T(r_{H})=T(\infty)=\infty$
according to the formula (11).
Thus, the function $T(r)$ must possess a minimum at some finite radius $r=r_{extrem}$.
At the radius, the equations (10) and (13) hold,
which means $r=r_{extrem}$ corresponds to the position of
null circular geodesics. In a word, we prove
the existence of null circular geodesics outside
extremal black holes.

\section{Conclusions}

We investigated on the existence of null circular geodesics
outside extremal spherically symmetric asymptotically flat hairy black holes.
We firstly obtained the null circular geodesic characteristic
equations. Then we got the equation of the fastest circular trajectory
that particles can circle
the central extremal spherically symmetric asymptotically flat hairy black hole.
We showed that the null circular geodesic equation
is identical to the fastest circular trajectory equation.
So the extreme period circular radius equation and
the null circular geodesic equation share the same roots.
We found that the extreme period circular radius equation possesses solutions.
So we proved that the null circular geodesic exists outside extremal spherically symmetric
asymptotically flat hairy black holes. It is natural that our proof also
holds for non-extremal spherically symmetric
asymptotically flat hairy black holes.

\begin{acknowledgments}

This work was supported by the Shandong Provincial Natural Science Foundation of China under Grant
No. ZR2022MA074. This work was also supported by a grant from Qufu Normal University
of China under Grant No. xkjjc201906, the Youth Innovations and Talents Project of Shandong
Provincial Colleges and Universities (Grant no. 201909118), Taishan Scholar Project of Shandong Province (Grant No.tsqn202103062)
and the Higher Educational Youth Innovation Science
and Technology Program Shandong Province (Grant No. 2020KJJ004).

\end{acknowledgments}

\end{document}